\documentclass[aps,prl,twocolumn]{revtex4-1}
\usepackage{amssymb}
\usepackage{amsmath,bm}
\usepackage{graphicx,color}
\usepackage{bbm}
\usepackage{multirow}
\newcommand{\B}[1]{{\bm{#1}}}

\usepackage{hyperref}

\begin{document}

\title{Transition from Static to Dynamic Friction in an Array of Frictional Disks}
\author{Harish Charan$^1$, Joyjit Chattoraj$^2$ Massimo Pica Ciamarra$^2$ and Itamar Procaccia$^1$ }
\affiliation{$^1$Department of Chemical Physics, the Weizmann Institute of Science,
Rehovot 76100, Israel.\\
$^2$ School of Physical and Mathematical Sciences,
Nanyang Technological University, Singapore}

\begin{abstract}
The nature of an instability that controls the transition from static
to dynamical friction is studied in the the context of an array of
frictional disks that are pressed from above on a substrate. In this case
the forces are all explicit and Newtonian
dynamics can be employed without any phenomenological assumptions.
We show that an oscillatory instability that had been discovered recently
is responsible for the transition, allowing individual disks to spontaneously
reach the Coulomb limit and slide with dynamic friction. The transparency of
the model allows a full understanding of the phenomenon, including the speeds of the waves
that travel from the trailing to the leading edge and vice versa.
\end{abstract}

\maketitle

{\bf Introduction}: The transition from static to dynamical friction is an old problem that intrigued already the ancient Greeks \cite{14Sam}. Explicitly, Themistius stated in 350 A.D. that ``it is easier to further the motion of a moving body than to move a body at rest". The phenomenon is central to many different fields of physics and material science including tribology, performance of micro electromechanical systems, mechanics of fracture,
and earthquakes. Despite the considerable amount of work in the modern era starting with Leonardo, Amonton and Coulomb \cite{15PP}, and culminating with enlightening experiments and simulations in recent years \cite{04RCF,09BBU,10BCF,18Sahli,13VMUZT,13BSBB,17BTNSMZ},
the actual instability mechanism that results in this transition is still debated \cite{13BSBB,11TSATM}. The aim of this Letter is to offer a very simple model for which the prevailing instability mechanism can be understood in full detail. This is done with the conviction that simple models that can be fully understood play an important role in statistical and nonlinear physics where they can often shed light on complex phenomena that may exhibit universal characteristics. Examples are the Ising model for magnetic phase transitions \cite{71Sta} or one-dimensional maps for the onset of chaos \cite{78Fei}. We thus do not attempt to model a specific physical realization, but consider a model that displays the desired transition whose triggering instability and its consequences are manifest and fully understood.

\begin{figure}
    \includegraphics[width=0.40\textwidth]{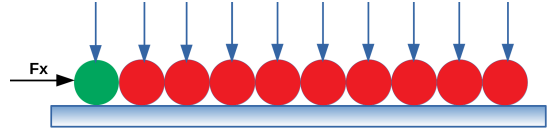}
    \caption{The model consists of $N$ identical disks (here and in the simulations below $N=10$) which interact via
    Hertz and Mindlin forces between themselves and the substrate below. A constant force $F_y$ is applied
    to press them against the substrate, and an external force $F_x$ is applied to the first disk, incresing it
    quasistatically until a pair of complex eigenvalues gets born, signalling an oscillatory instability. From that point on the Newtonian dynamics takes the system from static to dynamical friction. }
    \label{model}
    \end{figure}

The model is shown in Fig.~\ref{model}. It consists of $N$ 2-dimensional disks of radius R, with their initial center of mass positioned at $x_i=(2i -1)R; y_i=R$, $i=1\cdots N$, aligned over an infinite substrate at $y=0$. Each disk is pressed with an identical force $F_y$ normal to the substrate, providing a very simple model of asperities contacts in more realistic systems. Note that the substrate is totally
flat, in contrast to many attempts to explain the desired transition with periodic substrate, see for example \cite{96BFH}. The boundary conditions are periodic such that
the disk $i=N$ is in contact with the disk $i=1$. The forces and torques are annulled by force minimization protocol (see Supplementary Material) to reach mechanical equilibrium. After attaining equilibrium we increase quasistatically a force $F_x$ which is applied at the center of mass of the disk $i=1$. At some critical value of $F_x$ (and see below for details) the system becomes
unstable with respect to an oscillatory instability \cite{19CGPPa,19CGPPb}. This instability can trigger a transition from static to dynamical friction which is the subject of this Letter.
The instability will be shown to result from the fully Newtonian dynamics of the model, employing standard forces as discussed next, requiring no phenomenological input beyond the definition of the forces. The transition can
be seen in a movie provided at \href{https://www.dropbox.com/s/uiiu762jrkz7zga/Stattic-to-Dynamic-Movie.mp4?dl=0}{Movie-Link}.

{\bf The forces}: The interaction forces between the disks and between the disks and the substrate are standard \cite{79CS}. The normal force is determined by the overlap $\delta_{ij} \equiv 2R-r_{ij}$ between the disks and the compression $\delta_{iw} \equiv R-y_i$ between the disk and the substrate wall, $\B r_{ij}\equiv \B r_i-\B r_j$. We choose a Hertzian force between the disks,
\begin{equation}
\B F_{ij}^{(n)} = k_n \delta_{ij}^{3/2}\hat r_{ij} \ , \quad \hat r_{ij} \equiv \B r_{ij}/r_{ij} \ ,
\label{Fn}
\end{equation}
and similarly for the normal force between a disk and the substrate.
The reader should note that this force is
a model force, and although its use is time honored, it should be considered as an effective
force since we do not specify the precise deformation of the contact area between the disks and between the disks and the substrate.
The tangential force is a function of the tangential displacement $\B t_{ij}$ between the disks and $\B t_{iw}$ between the disks and the substrate. Upon first contact between the disks and the substrate, $t_{ij}= t_{iw}=0$. Providing every particle with the angular coordinate $\B\theta_i$, the change in tangential displacement is given by
\begin{eqnarray}
d\B t_{ij} &=&d\B r_{ij} -(d\B r_{ij}\cdot \B r_{ij}) \hat r_{ij} +\hat r_{ij} \times (R d\B\theta_i +R d\B\theta_j) \nonumber\\
d\B t_{iw} &=&(dx_i+R ~ d\theta_i)\hat x\ ,
\end{eqnarray}
where $\hat x$ is the unit vector parallel to the substrate.
For the tangential force we choose
the Mindlin model \cite{49Min} which between the disks reads
\begin{equation}
\B F_{ij}^{(t)} = -k_t\delta_{ij}^{1/2}t_{ij} \hat t_{ij} \ ,
\label{Min}
\end{equation}
with a similar definition for the tangential force between the disk and the substrate.
The tangential forces satisfy the Coulomb condition
\begin{equation}
\B F_{ij}^{(t)} \le \mu \B F_{ij}^{(n)} \ ,
\label{Coul}
\end{equation}
where $\mu=10$ is the friction coefficient.
For technical purpose we smooth out the Coulomb law such that the tangential force will have smooth derivatives; we choose:
\begin{equation}
\!\!\B F_{ij}^{(t)}\! =\! -k_t\delta_{ij}^{1/2}\!\left[1\!+\!\frac{t_{ij}}{t^*_{ij}} \!-\!\left(\frac{t_{ij}}{t^*_{ij}}\right)^2\right]\!t_{ij} \!\hat t_{ij} \ , ~
t^*_{ij}\! \equiv\! \mu \frac{k_n}{k_t} \delta_{ij} \ .
\label{Ft}
\end{equation}
Now the derivative of the force with respect to $t_{ij}$ vanishes smoothly at $t_{ij}=t^*_{ij}$ and Eq.~(\ref{Coul}) is fulfilled.

{\bf Dynamics and transition}: The dynamics is Newtonian; denoting the set of coordinates $\B q_i=\{\B r_i, \B \theta_i\}$:
\begin{eqnarray}
   \label{Newton2a}
   m\frac{d^2 \B r_{i}}{dt^2}&=&{\B F}_i(\B q_{i-1},\B q_i,\B q_{i+1})\ , \quad q_{N+1} = q_1 \ ,\\
   \label{Newton2b}
   I\frac{d^2 \B \theta_{i}}{dt^2}&=&{\B T}_i(\B q_{i-1},\B q_i,\B q_{i+1})\ ,
\end{eqnarray}
where $m$ and $I$ are the mass and moment of inertia of the $i$th disk, ${\B F}_i$ is the total force on disk $i$ and ${\B T}_i$ is the torque on that disk. Below time is measured in units
of $\sqrt{m2Rk_n}$ and length in units of $2R$. Although the simulations were performed for a wide range of parameters, we present results for $k_n=2000$, $k_t=2k_n/7$, $F_y=0.1$. The conditions for mechanical equilibrium
are ${\B F}_i= {\B T}_i=0$, and the stability of an equilibrium state is determined by the Jacobian matrix \begin{equation}
J_{ij}^{\alpha\beta} \equiv \frac{\partial \tilde F^\alpha_i}{\partial q_j^\beta}\ , \quad \tilde{ \B F}_i \equiv \sum_j \tilde{\B F}_{ij} \ ,
\end{equation}
where $\B q_j$ stands for either a spatial position or a tangential coordinate, and $\tilde {\B F}_i$ stands for either a force or a torque. The explicit calculation of the Jacobian matrix for the present model is presented
in the Supplementary Material. For our purpose here it is enough to note that the Jacobian matrix is real
but not symmetric, and therefore it can have pairs of complex eigenvalues which necessarily lead to an
oscillatory instability \cite{19CGPPa,19CGPPb}. Even if initially all the eigenvalues are real and the system is stable,
by increasing the external force $F_x$ we always reach a threshold value of this force where a pair
of complex eigenvalues is born. Details of this protocol are in the Supplementary Material. 

\begin{figure}
    \includegraphics[width=0.40\textwidth]{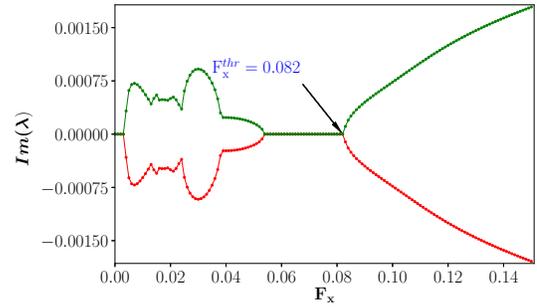}
    \caption{Two examples of a bifurcation of a pair of imaginary parts of eigenvalues upon increasing the external force $F_x$. A pair of real eigenvalues coalesces when the pair of imaginary eigenvalues bifurcates. The first bifurcation results in an oscillatory instability but the force $F_x$ is not sufficient to trigger a transition to dynamical friction. The second does.}
    \label{birth}
    \end{figure}

\begin{figure}
\includegraphics[width=0.40\textwidth]{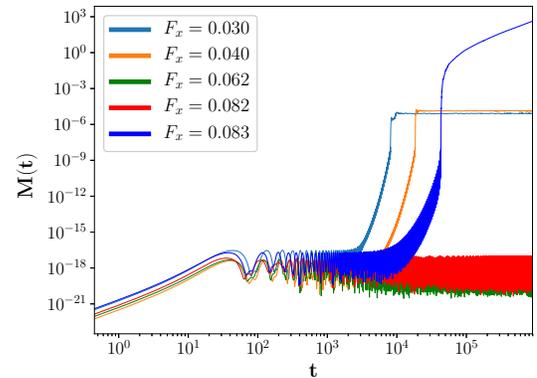}
    \caption{Dynamical response of $M(t)$ at various values of $F_x$. The two lowest values are within the
    range of existence of an oscillatory instability (see Fig.~\ref{birth}), but $F_x$ is not large enough to trigger a transition to dynamical friction. For $F_x=0.062$ the system is stable and $M(t)$ remains
    minute. The largest value of $F_x=0.083$ is after the birth of the new pair of complex eigenvalue, and now
    the oscillatory instability triggers the transition to dynamical scaling.}
    \label{MSD}
 \end{figure}

{\bf Instability and Dynamics}. The birth of the instability is demonstrated in Fig~\ref{birth}. Note that this figure
exhibits two events of birth of a complex conjugate pair of eigenvalues. The first complex pair dies out with increasing $F_x$ without triggering a transition to dynamical friction. The difference in dynamical response is presented in Fig.~\ref{MSD} which shows the mean-square displacement (MSD) $M(t)=\sum_{i=1}^{10} |\B r_i(t)-\B r_i(0)|^2$. The protocol resulting in these dynamics is spelled out in detail in the Supplementary Material. 
 \begin{figure}
\includegraphics[width=0.33\textwidth]{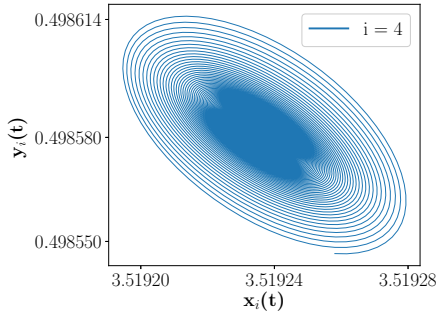}
    \caption{The actual trajectory in the $x-y$ plane (for $t\approx 17500$) that is induced by the 
    instability. Here we show the fourth grain ($i=4$) but this is representative for all grains. }
    \label{spiral}
    \end{figure}
 For the two external forces $F_x=0.03$ and 0.04 the oscillatory instability exists but the MSD becomes
 stationary at a minute fraction the disk radius (about $10^{-5}$), indicating static friction with charging of $t_{iw}$ but without transition to dynamical friction. For $F_x=0.062$ all the eigenvalues of $\B J$ are real,
 there is no instability, and $M(t)$ never increases. For $F_x=0.083$ the system is in the second regime
 of oscillatory instability, which is now developing strongly, bringing $M(t)$ to about $10^{-2} R$ where
 for the first time the Coulomb law is breached at the contact with the substrate, and subsequently $M(t)$ increases linearly in time with a fixed average velocity.

\begin{figure}
\includegraphics[width=0.37\textwidth]{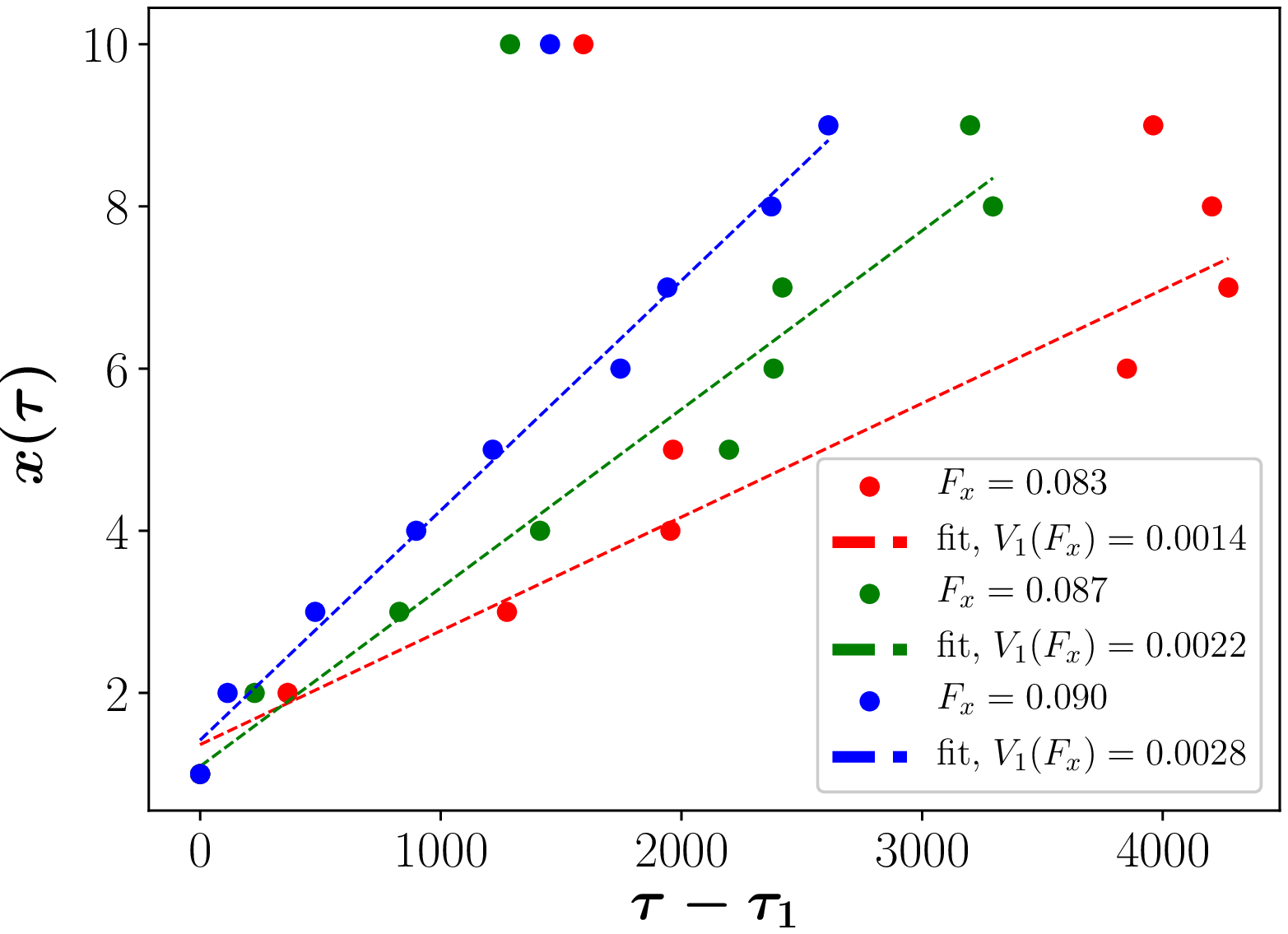}
\includegraphics[width=0.37\textwidth]{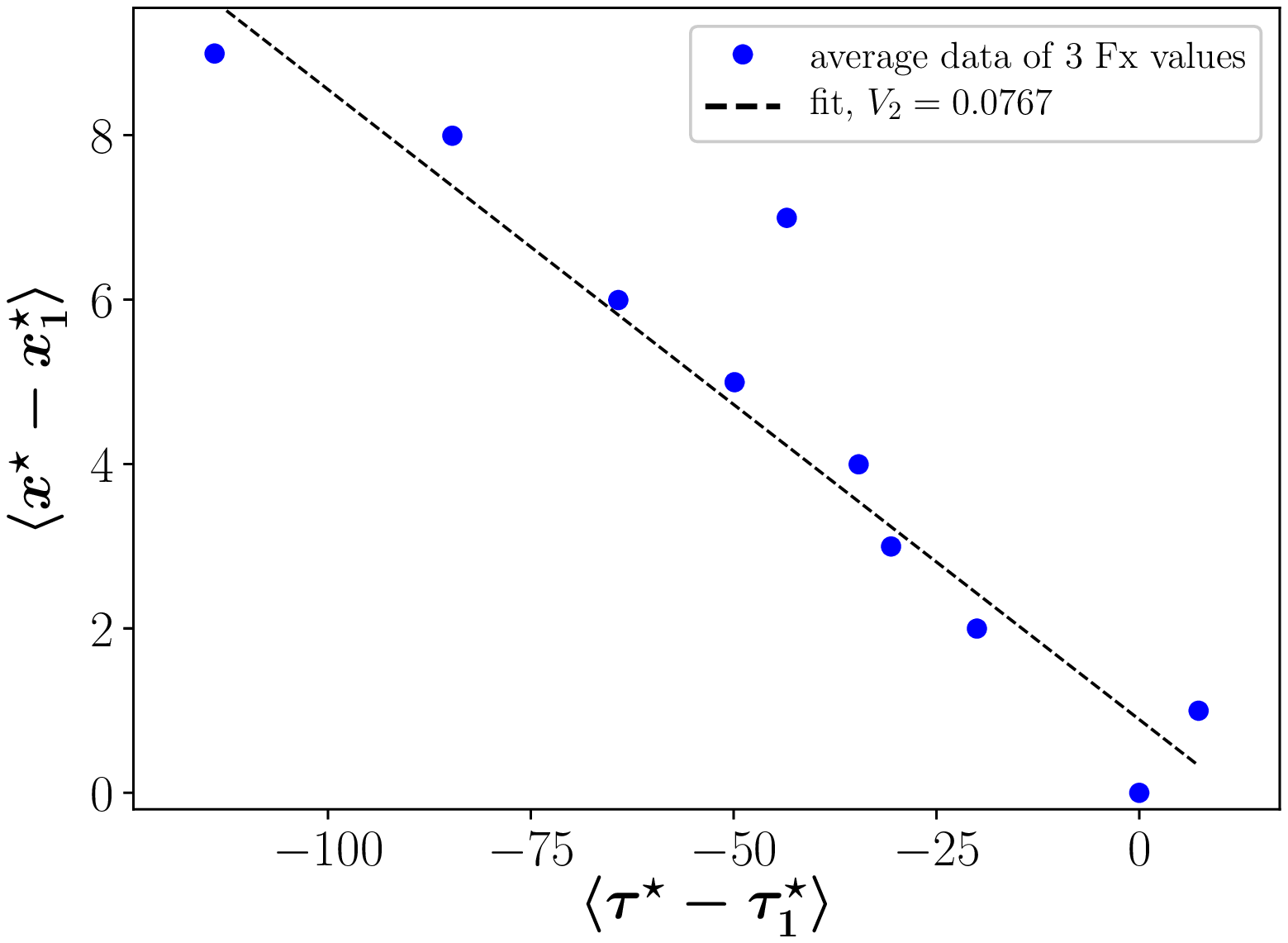}
\caption{Upper panel: the wave of breaching of the Coulomb law, starting with the first
disk and ending with the last. Every realization is noisy, but averaging on 25 realizations
yields this plot which supports a wave velocity $V_1(F_x)$ that depends on the external force $F_x$. Lower panel: The wave of actual sliding, starting with the last disk and ending with the first. This backward running
wave has a velocity $V_2$ that does not depend on the external force $F_x$.}
\label{waves}
\end{figure}

 The mechanism for the transition is understood once we examine the typical trajectories that the centers of
 mass of disks follow as a result of the instability. An example pertaining to the fourth disk is shown in Fig.~\ref{spiral}, and is representative for all the other disks. A spiral in the $x-y$ plane leads  in time to an increase in $y_i$ to the point that breaching the Coulomb law at the contact with the substrate becomes inevitable. In consequence the tangential forces at contact fail to oppose the external force $F_x$ and a transition to dynamical friction can take place. The actual onset of dynamic friction is in close accord with experimental and simulational reports \cite{04RCF,10BCF,13BSBB}. The first disk reaches the Coulomb limit at some time $\tau_1$, and consequently a forward wave of such events runs through the system, with the second, third, fourth etc. disk reaching the Coulomb limit at times $\tau_i$ with $\tau_{i+1}> \tau_i>\tau_1$ and
 $i$ being the index of the disk. While each realization is somewhat stochastic, averaging over 25 realizations
 as shown in Fig.~\ref{waves} indicates that
 \begin{equation}
V_1(F_x)\times (\tau_i-\tau_1)\approx x_i ,
 \label{firstv}
 \end{equation}
 with a wave speed $V_1(F_x)$ depending on the external force $F_x$. In Fig. \ref{waves} upper panel $\tau$ denotes the time that a disk breaches the Coulomb low for the first time and $x(\tau)$ is the position of the disk when this takes place.   For $F_x=0.083, 0.087, 0.090$ the corresponding velocity of this front is $0.0014\pm 0.00013$, $0.0022\pm 0.00015$ and $0.0028\pm 8\times 10^{-5}$.
 These velocities are explained below. 
 
 The actual sliding begins at the leading (last) disk, with a backward wave of sliding traveling from the
 last to first disk. This is also in accord with experimental and simulational observations \cite{10BCF,13BSBB,13BSBB}. The speed of the backward wave
 is independent of $F_x$, indicating that it is an inherent elastic wave, as is explained below.  It is also in agreement
 with observations that the speed $V_2$ of the backward wave of sliding is much faster than $V_1$ \cite{04RCF}. In the lower panel of Fig.~\ref{waves} the times $\tau^*$ are measured backwards from the first sliding of the last disk, and $x^*$
 is the position where the sliding occurs.  Since the velocity of the backward wave is independent of $F_x$ the lower panel of Fig.~\ref{waves} is obtained
 by averaging the data from simulations at the same three values of $F_x$ as the ones shown in the upper panel.

\begin{figure}
	\includegraphics[width=0.32\textwidth]{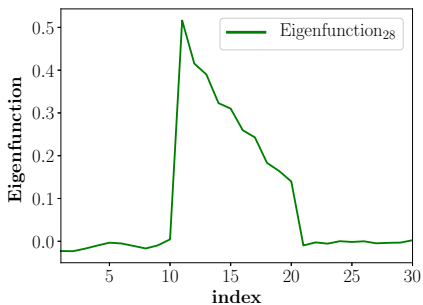}
	\caption{Projection of the shear-wave eigenfunction of the Jacobian matrix on the degrees of freedom $q_\ell$, 
		$x$ modes for $\ell=1$ to 10, $y$ modes for $\ell=11$ to 20, and angular modes for $\ell=21$ to 30. This mode
		is predominantly a $y$ shear
		wave that is responsible for initiating the sliding motion. The wave-length of this
		up down shear wave is approximately 2.}
	\label{eigen}
\end{figure}

{\bf Theory}: To understand the wave speeds we discuss first the backward wave of detachment which is dominated by 
an elastic transverse wave of up-down motion in the $y$ direction. Examining the 30 eigenfunctions of the Jacobian matrix $\B J$ one identifies only one 
which consists of a pure $y$ transverse wave, see Fig.~\ref{eigen}. This eigenfunction has no projection on either longitudinal
or angular degrees of freedom. It has an eigen-frequency $\omega\approx 0.235$ and a wavelength 
of approximately 2. Thus the associated $k$ vector is $k\approx 2\pi/2\approx \pi$, determining
a velocity $V_2\approx 0.235/\pi\approx 0.075$ in excellent agreement with the measured value of the
backward velocity. 

The forward wave of Coulomb breaching is not related to the internal elastic waves in the system. Rather, it
is related to the imaginary part of the pair of complex eigenvalues, which governs the exponential growth rate of the spiral trajectories and therefore the push by the $i$th on the ($i+1$)th disk, resulting in breaching the Coulomb threshold. Indeed, while the backward wave-speed is basically independent of $F_x$, the forward wave-speed must be a function of $F_x$ since the imaginary part of the eigenvalues grows with increasing $F_x$. For the three values of $F_x$ shown in the upper panel of 
Fig.~\ref{waves} the imaginary part of the associated eigenvalue is $0.00024, 0.00077$ and $0.001$ respectively. Using the values of the velocities measured for the three different values of $F_x$ we compute the ratios of these velocities to the growth rates to find an approximate constant value  $4.3\pm 1.5$. 
 \begin{figure}
\includegraphics[width=0.37\textwidth]{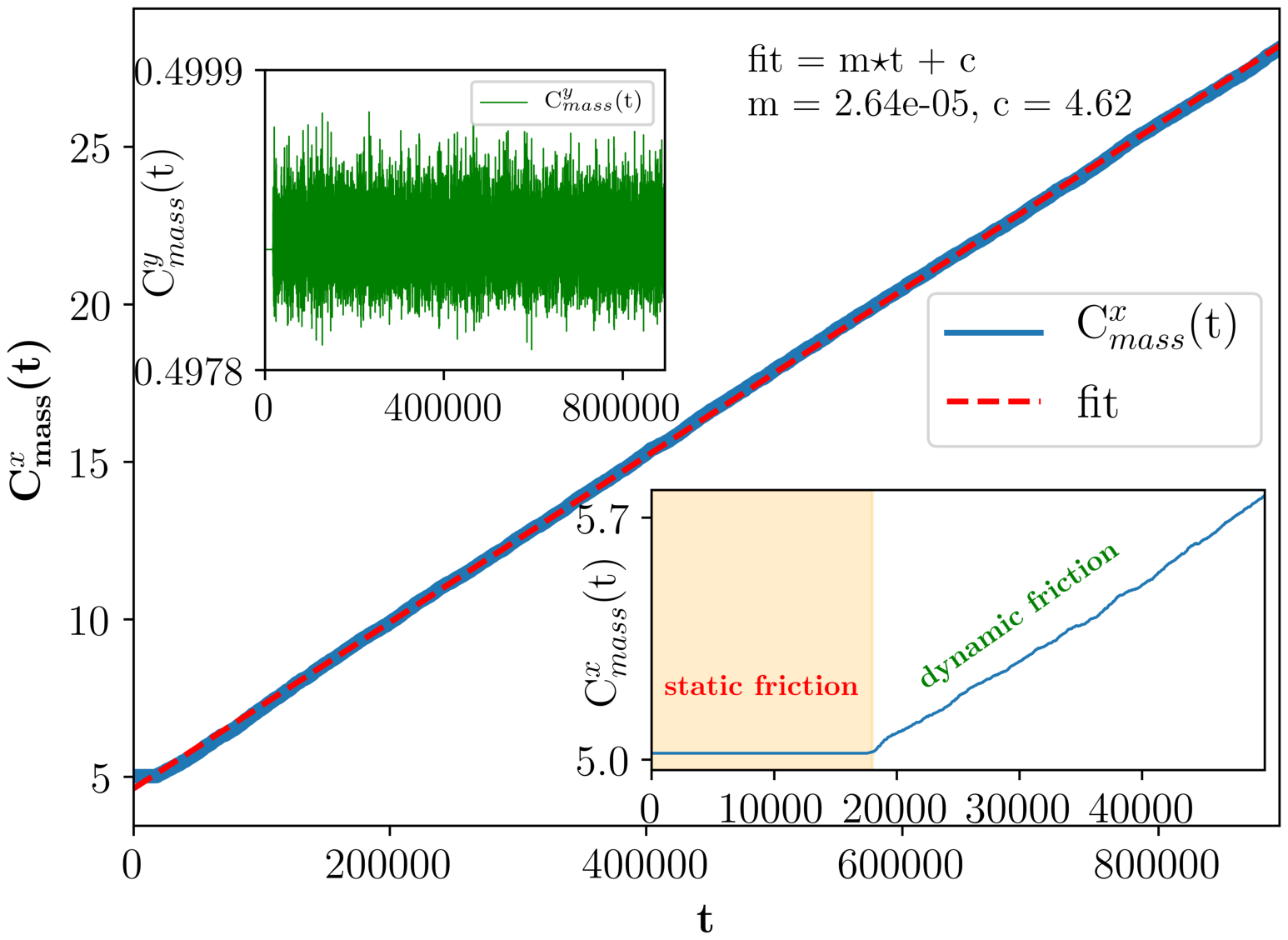}
\includegraphics[width=0.37\textwidth]{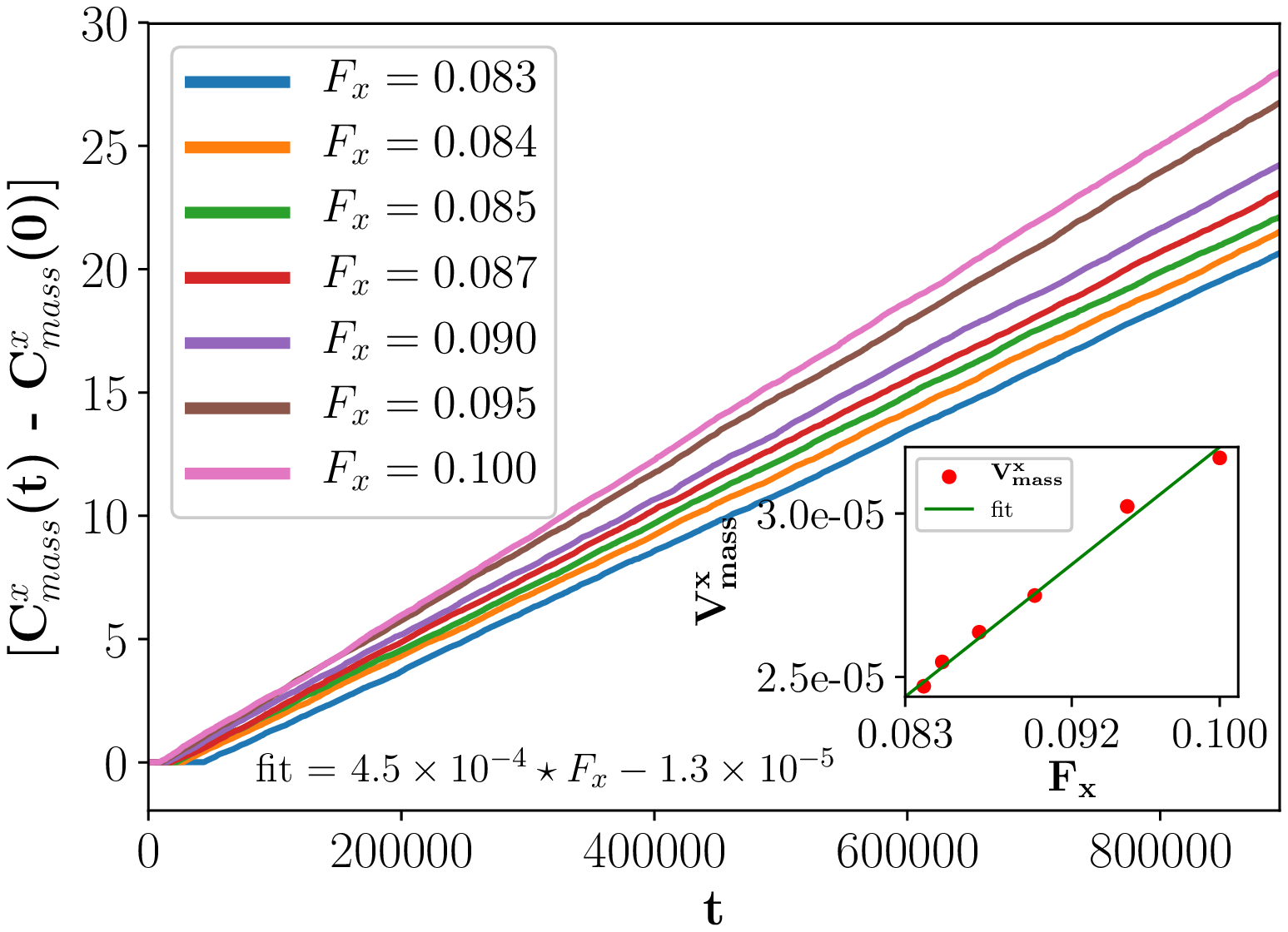}
\caption{The transition from static to dynamical friction as measured by the dynamics of the center of mass.
Upper panel: the motion of the $x$ component of the center of mass for $F_x=0.087$. The upper inset shows
the $y$ component of the center of mass that oscillates without growth. The lower inset shows the transition 
itself from static to dynamics. Lower panel: the dynamics of the $x$ component of the center of mass for
different values of $F_x$. The inset shows how the velocity of the center of mass depends on $F_x$. }
\label{dyn.friction}
\end{figure}

The transition as seen from the dynamics of the center-of-mass of the whole system is shown
in the upper panel of Fig.~\ref{dyn.friction}. Here the $x$ and $y$ components of the center
of mass are defined as $C^x_{\rm mass}\equiv \frac{1}{N}\sum_i^N x_i$ and equivalently for $y$.
In the upper inset we show $C^y_{\rm mass}$ which is oscillating without growth. The main figure
displays $C^x_{\rm mass}$ as a function of time. The lower
inset shows the transition itself, which on the scale shown appears like a simple
first order transition from static to dynamics with a fixed velocity $V^x_{\rm mass}$, hiding all the microscopic richness discussed above.
In the lower panel we display the dependence of the constant velocity in the dynamic friction regime
on the external force $F_x$. The inset indicates a liner relation between the two, with the best fit reading
\begin{equation}
V_{\rm mass}^x \approx -1.3\times 10^{-3} + 4.5\times 10^{-4} F_x \ .
\label{VxFx}
\end{equation}

{\bf Summary}: The rich physics associated with the transition from static to dynamical friction is demonstrated 
using a very simple model of $N$ disks on a flat substrate, interacting via the often used Hertz and Mindlin
forces. By increasing the external force $F_x$, Newtonian dynamics without further phenomenological assumptions result
in an oscillatory instability that ends up with the observed transition. In agreement with much more complicated
experimental and simulational examples, the transition is associated with forward and backward running waves of 
events that accompany the transition. An advantage of the simple model is that the instability, its development, and the
wave speeds observed can be all understood in full detail.  It should be of utmost interest to examine the mechanism 
described here also in experimental contexts.

{\bf Acknowledgments:}
This work has been supported by the ISF-Singapore program and by the US-Israel BSF.
We thank Eran Bouchbinder and Yuri Lubomirsky 
for useful discussions regarding the wave propagations.
JC and MPC were suppoted by the Singapore Ministry of Education
through the Academic Research Fund (Tier 2) MOE2017-T2-1-066 (S), and the NSCC for computational resources.

\end{document}


\title{Supplementary Material to ``Transition from Static to Dynamic Friction in an Array of Frictional Disks"}
\author{Harish Charan$^1$, Joyjit Chattoraj$^2$, Massimo Pica Ciamarra$^2$ and Itamar Procaccia$^1$ }
\affiliation{$^1$Department of Chemical Physics, the Weizmann Institute of Science,
	Rehovot 76100, Israel.\\
	$^2$ School of Physical and Mathematical Sciences,
	Nanyang Technological University, Singapore}
\thispagestyle{empty}		
\maketitle			
{\bf The contents of the this supplementary material}: We first describe below the computational protocol leading to
Figs.~2 and 3 in the main text. Secondly we present the analytic calculation of the Jacobian matrix for the model described
in the main text. 

{\bf Computational protocol for preparing a mechanically equilibrated system}:
In the simulations we incorporate ``over-damped'' and ``Newtonian'' algorithms \cite{19CGPPa,19CGPPb}. The ``over-damped'' 
algorithm incorporates a viscous drag force to the equations of motion to force the system to reach equilibrium with
zero force and torque on each disk. The global viscous damping force is proportional to 
the disk velocities with the proportionality constant equals to $m\nu$ with $\nu = 0.01$.  The ``Newtonian'' algorithm  
evolve the system under Newton's equation of motions without any drag forces. Each ``Over-damped'' run is 
performed in two stages; in the first state only the external force $F_y$ is applied and the system is equilibrated.  
In the second stage the external force $F_x$ is added and the system is equilibrated again. Finally the Newtonian dynamics
is allowed to develop under the external forces but without damping. 

We start with a chain of identical disks placed on a horizontal flat substrate. The disks are then 
pushed down by an external force $\vec{F}_{y}$ in vertically downward direction[here the horizontal 
direction is  $\hat{x}$ and vertical direction is $\hat{y}$]. Over-damped dynamics is used to reach 
a mechanical equilibrium state.  The system is then acted upon a series of 
increasing force $\vec{F}_{x}$ in a quasi-static fashion, meaning that we annul the forces and the torques before 
increasing the value of $\vec{F}_{x}$.  At this state, we calculate the Jacobian matrix of the 
system and its eigenvalues. 

At a given value of $\vec{F}_{x}$ one observes the birth of a pair of complex eigenvalues. 
The value of the imaginary part of this complex eigenvalue increases with 
$F_{x}$ as shown in Fig.~2 of the main article.
For the current study, discrete element method has been implemented by employing LAMMPS code \cite{Plimpton1995}. 

The configuration at the end of Stage-2 is evolved  using Newton's equations of motion
with no damping for all the values of the force $F_x$ for which complex eigenvalues exist. 
For the present parameters in the simulation the system becomes unstable for all force 
values $F_{x} > 0.082$, triggering the oscillatory instability which then leads eventually to the transition
from static to dynamical friction.

{\bf Derivation of the Jacobian matrix}: Newton's equation of the motion for a system of disks is given by:
\begin{equation}\label{eq1}
\textbf{M}\ddot{\textbf{Q}} = \textbf{F(\textbf{Q})}
\end{equation}
Here \textbf{Q} is the vector of positional and rotational coordinates, \textbf{M} is the vector of masses and moments of inertia of the individual disks.
In a two-dimensional ($xy$) system of N disks the moment of inertia is 
denoted as $\textnormal{I}_{k}$ and the corresponding torque is denoted as T$_{k}$. Then each term in the
above equation can be written in the matrix form as follow:
\begin{widetext}
\begin{equation}\label{eq2}
\bf{F(Q)} = 
\begin{bmatrix}
f_{1}^{x}\\
\vdots\\
f_{N}^{x}\\
f_{1}^{y}\\
\vdots\\
f_{N}^{y}\\
\textnormal{T}_{1}\\
\vdots\\
\textnormal{T}_{N}\\
\end{bmatrix};
\,\,
\bf{Q} = 
\begin{bmatrix}
r_{1}^{x}\\
\vdots\\
r_{N}^{x}\\
r_{1}^{y}\\
\vdots\\
r_{N}^{y}\\
\theta_{1}\\
\vdots\\
\theta_{N}\\
\end{bmatrix}
\,\textnormal{and}\,
\bf{M} = 
\begin{bmatrix}
m_{1}  & 0       & \cdots  & \cdots & \cdots & \cdots & \cdots & 0 \\
0      & \ddots  & \ddots  & \ddots & \ddots & \ddots & \ddots & \vdots \\
\vdots & \ddots  & \ddots  & \ddots & \ddots & \ddots & \ddots & \vdots \\
\vdots & \ddots  & \ddots  & \textnormal{m}_{N} & \ddots & \ddots & \ddots & \vdots \\
\vdots & 0       & \ddots  & \ddots & \textnormal{I}_{1}  &  \ddots & \ddots & \vdots \\
\vdots & \ddots  & \ddots  & \ddots & \ddots & \ddots & \ddots & \vdots \\
\vdots & \ddots  & \ddots  & \ddots & \ddots & \ddots & \ddots & \vdots \\
0 & \cdots &  \cdots & \cdots & 0 &  \cdots & \cdots & \textnormal{I}_{N}\\
\end{bmatrix}
\end{equation}
Now after Taylor expanding  Eq.\cref{eq1} around equilibrium \textbf{Q} = {\bf Q$_{{\bf 0}}$}, we get
\begin{equation}\label{eq3}
\textbf{M}(\ddot{\textbf{Q}} - \ddot{\textbf{Q}}_{\textbf{0}}) = \textbf{F}(\textbf{Q} - \textbf{Q}_{\textbf{0}}) =
\underbrace{\textbf{F}(\textbf{Q}_{\textbf{0}})}_\text{0} +  
{\bf F^\prime}(\textbf{Q})\Big|_{\textbf{Q}_{\textbf{0}}}(\textbf{Q} - \textbf{Q}_{\textbf{0}})
+ \frac{1}{2!}\underbrace{{\bf F^{\prime \prime}}(\textbf{Q})\Big|_{\textbf{Q}_{\textbf{0}}}(\textbf{Q} - \textbf{Q}_{\textbf{0}})^{2}}_\text{neglecting} + \cdots
\end{equation}
which leads to
\begin{equation}\label{eq4}
\textbf{M} \ddot{{\bf \Delta}} = \B J{\bf \Delta}
\end{equation}
\end{widetext}
where, ${\bf \Delta} =  \textbf{Q} - \textbf{Q}_{\textbf{0}}$,
$\B J = \Big|\Big|\pdv{\textbf{F}_{i}}{\textbf{Q}_{j}}\Big|\Big|_{\bf Q_{{\bf 0}}}$. 
The stability of a system is determined by the Jacobian matrix $\B J$. In the following sections of this 
supplementary-material, we provide the Jacobian for the model discussed in the main text.

{\bf Tangential displacements}: The normal force laws were described in detail in the main text. Here we
spell out the calculation of the tangential forces. 
The tangential displacements can be divided into two parts: 
(1) disk-disk interaction and (2) disk-wall interaction

{\bf Disk-disk pair}: The tangential velocity which is the derivative of the tangential displacement {\bf t}$_{ij}$ of a pair 
of interacting disks $i$ and $j$ is given by:
\begin{equation}\label{eq11}
{\bf v}_{ij}^{t} = \frac{\textrm{d}{\bf t}_{ij}}{\textrm{d}t} = {\bf v}_{ij} - {\bf v}^{n}_{ij} + 
\hat{r}_{ij} \times (\textnormal{R}_{i}\boldsymbol{\omega}_{i} + \textnormal{R}_{j}\boldsymbol{\omega}_{j})
\end{equation}
for disks of equal radii, $\textnormal{R}_{i} = \textnormal{R}_{j} = \textnormal{R}$ 
\begin{equation}\label{eq12}
\frac{\textrm{d}{\bf t}_{ij}}{\textnormal{d}t} = {\bf v}_{ij} - {\bf v}^{n}_{ij} + 
\hat{r}_{ij} \times \textnormal{R}(\boldsymbol{\omega}_{i} + \boldsymbol{\omega}_{j})
\end{equation}
where, ${\bf v}_{ij} = {\bf v}_{i} - {\bf v}_{j}$ is the relative velocity of the pair-$i$ and $j$. 
${\bf v}_{ij}^{n}$ is the projection of the ${\bf v}_{ij}$ along the direction $\hat{r}_{ij}$.
$\boldsymbol{\omega}_{i}$ and $\boldsymbol{\omega}_{j}$ are the angular velocities of the disks $i$ 
and $j$ respectively given by $\boldsymbol{\omega}_{i/j} = \frac{\textrm{d}\boldsymbol{\theta}_{i/j}}{\textrm{d}t}$,
here, $\textrm{d}\boldsymbol{\theta}_{i/j}$ is the angular displacement of disk $i/j$.
The above equation can be written in differential form as follow:
\begin{equation}\label{eq13}
\textrm{d}{\bf t}_{ij} = \textrm{d}{\bf r}_{ij} - \textrm{d}{\bf r}^{n}_{ij} + 
\hat{r}_{ij} \times R(\textrm{d}\boldsymbol{\theta}_{i} + \textrm{d}\boldsymbol{\theta}_{j})
\end{equation}
For a quasi-two-dimensional system as is in our studies, $\boldsymbol{\omega}_{i}$ and 
$\boldsymbol{\theta}_{i}$ will have only one component in $\hat{z}$-direction perpendicular to the 
$xy$-plane, therefore, 
$\hat{r}_{ij} \times \textnormal{R}\textrm{d}\boldsymbol{\theta}_{i} = 
R\textrm{d}\theta_{i}(y_{ij}\hat{x} - x_{ij}\hat{y})/r_{ij}$. Hence, Eq.(\ref{eq13}) can be written in 
tensorial form as follow:
\begin{equation}\label{eq14}
\textrm{d}t^{\alpha}_{ij} = \textrm{d}r^{\alpha}_{ij} - (\textrm{d}{\bf r}_{ij}\cdot\hat{r}_{ij})\frac{r^{\alpha}_{ij}}{r_{ij}} 
+ (-1)^{\alpha}R(\textrm{d}\theta_{i}+\textrm{d}\theta_{j})\frac{r^{\beta}_{ij}}{r_{ij}} 
\end{equation}
where $\alpha$ and $\beta$ can take value 0 and 1 which correspond to $x$ and $y$ components, 
respectively. 

Angular displacement d$\theta_{i}$ of a disk-$i$ remains unaffected when disk-$i$ 
changes its transnational position, i.e. $\frac{\textrm{d}\theta_{j}}{\textrm{d}r^\alpha_{j}} = 0$.
Thus, the change in tangential displacement along $\beta$ due to the change in position of disk-$i$
along $\alpha$ only contributes in translations, and it can be written as (using Eq.(\ref{eq14}))
\begin{equation}\label{eq15}
\frac{\textrm{d}t^{\beta}_{ij}}{\textrm{d}r^{\alpha}_{j}} = 
-(\Delta_{\alpha\beta} - \frac{r^{\alpha}_{ij} r^{\beta}_{ij}}{r^{2}_{ij}})
\end{equation}
where $\Delta_{\alpha\beta}$ is the Kronecker delta which is one when $\alpha = \beta$, or else zero. 
Similarly, a change in rotational coordinates does not modify the particles relative distance, i.e.
$\frac{\textrm{d}r^{\beta}_{ij}}{\textrm{d}\theta_{j}} = 0$. Therefore, a change in tangential 
displacement along $\beta$ with respect to the change in $\theta_{j}$ reads (from Eq.(\ref{eq14}))
\begin{equation}\label{eq16}
\frac{\textrm{d}t^{\beta}_{ij}}{\textrm{d}\theta_{j}} = 
(-1)^{\beta}\textnormal{R}\frac{r^{\alpha}_{ij}}{r_{ij}}
\end{equation}
In the above equation, $\alpha$ and $\beta$ are always different. The magnitude of tangential 
distance $t_{ij}$ can be obtained from the relation $t^{2}_{ij} = \sum_{\alpha} t{^{\alpha}}^{2}_{ij}$. 
The differential of which follows 
$\textrm{d}t_{ij} = \sum_{\alpha} \frac{t^{\alpha}_{ij}}{t_{ij}}\textrm{d}t^{\alpha}_{ij}$.
Therefore, derivatives of magnitude of tangential distance $t_{ij}$ with respect to $r^{\alpha}_{j}$ 
and $\theta_{j}$ can be expressed as:
\begin{equation}\label{eq17}
\frac{\textnormal{d}t_{ij}}{\textnormal{d}r^{\alpha}_{j}} = 
\Bigg(\frac{t^{x}_{ij}}{t_{ij}}\Bigg)\frac{\textnormal{d}t^{x}_{ij}}{\textnormal{d}r^{\alpha}_{j}} +
\Bigg(\frac{t^{y}_{ij}}{t_{ij}}\Bigg)\frac{\textnormal{d}t^{y}_{ij}}{\textnormal{d}r^{\alpha}_{j}}
\end{equation}
\begin{equation}\label{eq18}
\frac{\textnormal{d}t_{ij}}{\textnormal{d}\theta_{j}} = 
\Bigg(\frac{t^{x}_{ij}}{t_{ij}}\Bigg)\frac{\textnormal{d}t^{x}_{ij}}{\textnormal{d}\theta_{j}} +
\Bigg(\frac{t^{y}_{ij}}{t_{ij}}\Bigg)\frac{\textnormal{d}t^{y}_{ij}}{\textnormal{d}\theta_{j}}
\end{equation}
The above differential equations can be solved by incorporating Eq.(\ref{eq15}) and Eq.(\ref{eq16}).
The derivative of the tangential threshold (since it is a linear function of overlap distance $\delta_{ij}$,
Eq.(5) of the main text with respect to $r^{\alpha}_{j}$ is given as:
\begin{equation}\label{eq19}
\frac{\textnormal{d}t^{\star}_{ij}}{\textnormal{d}r^{\alpha}_{j}} = 
\mu\Bigg(\frac{k_{n}}{k_{t}}\Bigg) \frac{r^{\alpha}_{ij}}{r_{ij}}
\end{equation}
which will be unaffected by the change in the rotation, i.e. 
$\frac{\textnormal{d}t^{\star}_{ij}}{\textnormal{d}\theta_{j}} = 0$ 

{\bf Disk-wall interaction} : In this subsection, we present the tangential displacement and its derivatives for the disk-wall ($iw$) 
interaction. For the disk-disk pair interaction, we performed the derivatives with respect to 
all the disk coordinates $r^{\alpha}_{j}$ and $\theta_{j}$, but for the disk-wall 
interaction we can differentiate with respect to the disk coordinates only. 
Thus Eqs.~(\ref{eq15}) and (\ref{eq19}) 
and hence Eq.(\ref{eq26}), Eq.(\ref{eq27}), Eq.(\ref{eq28}) and Eq.(\ref{eq32}) for disk-wall interaction will 
be modified only slightly. Following Eq.(\ref{eq14}) for the disk-disk pair, we 
can write the similar tangential displacement for the disk-wall interaction as follow:
\begin{equation}\label{eq20}
\textrm{d}t^{\alpha}_{iw} = \textrm{d}r^{\alpha}_{iw} - (\textrm{d}{\bf r}_{iw}\cdot\hat{r}_{iw})\frac{r^{\alpha}_{iw}}{r_{iw}} 
+ (-1)^{\alpha}R(\textrm{d}\theta_{i})\frac{r^{\beta}_{iw}}{r_{iw}} 
\end{equation}
Here for $x$, $\alpha$ and $\beta = 0$ and for $y$, $\alpha$ and $\beta = 1$. Note that there is no 
d$r_{w}$, or d$\theta_{w}$ as the wall remains always static.  
Since  we know that $\frac{\textrm{d}\theta_{i}}{\textrm{d}r^\alpha_{i}} = 0$, the above equation leads 
to:
\begin{equation}\label{eq21}
\frac{\textrm{d}t^{\beta}_{iw}}{\textrm{d}r^{\alpha}_{i}} = 
\Delta_{\alpha\beta} - \frac{r^{\alpha}_{iw} r^{\beta}_{iw}}{r^{2}_{iw}}
\end{equation}
similarly the derivative with respect to $\theta_{i}$ becomes:
\begin{equation}\label{eq22}
\frac{\textrm{d}t^{\beta}_{iw}}{\textrm{d}\theta_{i}} = 
(-1)^{\beta}\textnormal{R}\frac{r^{\alpha}_{iw}}{r_{iw}}
\end{equation}
In the above equation, $\alpha$ and $\beta$ are always different. Since for the disk-wall interaction
the tangential distance $t_{iw}$ can be obtained from the relation 
$t^{2}_{iw} = \sum_{\alpha} t{^{\alpha}}^{2}_{iw}$. 
The differential form of which reads as: 
$\textrm{d}t_{iw} = \sum_{\alpha} \frac{t^{\alpha}_{iw}}{t_{iw}}\textrm{d}t^{\alpha}_{iw}$.
This leads to the derivatives of magnitude of tangential distance $t_{iw}$ with respect to 
$r^{\alpha}_{i}$ and $\theta_{i}$ as follow:
\begin{equation}\label{eq23}
\frac{\textnormal{d}t_{iw}}{\textnormal{d}r^{\alpha}_{i}} = 
\Bigg(\frac{t^{x}_{iw}}{t_{iw}}\Bigg)\frac{\textnormal{d}t^{x}_{iw}}{\textnormal{d}r^{\alpha}_{i}} +
\Bigg(\frac{t^{y}_{iw}}{t_{iw}}\Bigg)\frac{\textnormal{d}t^{y}_{iw}}{\textnormal{d}r^{\alpha}_{i}}
\end{equation}
\begin{equation}\label{eq24}
\frac{\textnormal{d}t_{iw}}{\textnormal{d}\theta_{i}} = 
\Bigg(\frac{t^{x}_{iw}}{t_{iw}}\Bigg)\frac{\textnormal{d}t^{x}_{iw}}{\textnormal{d}\theta_{i}} +
\Bigg(\frac{t^{y}_{iw}}{t_{iw}}\Bigg)\frac{\textnormal{d}t^{y}_{iw}}{\textnormal{d}\theta_{i}}
\end{equation}
We already know that $t^{\star}_{iw} = \mu \frac{k_{n}}{k_{t}}\delta_{iw}$, therefore, the derivative 
of the this tangential threshold for the disk-wall interaction would follow as:
\begin{equation}\label{eq25}
\frac{\textnormal{d}t^{\star}_{iw}}{\textnormal{d}r^{\alpha}_{i}} = 
-\mu\Bigg(\frac{k_{n}}{k_{t}}\Bigg) \frac{r^{\alpha}_{iw}}{r_{iw}}
\end{equation}
and the quantity would be unaffected with respect to the change in the rotation of disk $i$, i.e. 
$\frac{\textnormal{d}t^{\star}_{iw}}{\textnormal{d}\theta_{i}} = 0.$
	
{\bf Evaluation of the J operator}:
The evaluation of the J operator is described in two section here. We first provide the calculations for 
the disk-disk pair which is followed by the disk-wall interaction. 

{\bf Disk-disk pair}: 
Derivative of the tangential force (Eq.(\ref{eq14})) with respect to $r^{\alpha}_{i}$:
\begin{widetext}
\begin{equation}\label{eq26}
\begin{array}{l}
\pdv{{\textnormal{F}^{t}_{ij}}^{\beta}}{r^{\alpha}_{j}} = 
-k_{t} \pdv{}{r^{\alpha}_{j}}\Bigg[\delta_{ij}^{1/2}\Bigg(t^{\beta}_{ij} + \tilde{t}t^{\beta}_{ij} - 
\tilde{t}^{2}t^{\beta}_{ij}\Bigg)\Bigg]\\
= \frac{1}{2} \delta^{-1}_{ij} \frac{r_{ij}^{\alpha}}{r_{ij}} {\textnormal{F}_{ij}^{t}}^{\beta} 
-\,\,k_{t}\delta^{1/2}_{ij}\Bigg[(1 + \tilde{t} - \tilde{t}^{2})\pdv{t^{\beta}_{ij}}{r^{\alpha}_{j}}
\,\,+\,\,(\tilde{t}^{\beta} - 2\tilde{t}\tilde{t}^{\beta})\pdv{t_{ij}}{r^{\alpha}_{j}} 
\,\,+\,\,(-\tilde{t}\tilde{t}^{\beta} + 2\tilde{t}^{2}\tilde{t}^{\beta})\pdv{t^{\star}_{ij}}{r^{\alpha}_{j}}\Bigg]
\end{array}
\end{equation}
\end{widetext}
Here, $\tilde{t} = \frac{t_{ij}}{t^{\star}_{ij}}$ and  
$\tilde{t}^{\beta} = \frac{t_{ij}^{\beta}}{t^{\star}_{ij}}$. 
The expressions for all the three partial differentiation in (Eq.(\ref{eq26})) are already shown in 
(Eq.(\ref{eq21})), (Eq.(\ref{eq23})), and (Eq.(\ref{eq25})).

Similarly, the derivative of tangential force with respect to $\theta_{j}$ (using the same notation as
above) can be found as:
\begin{equation}\label{eq27}
\pdv{{\textnormal{F}^{t}_{ij}}^{\beta}}{\theta_{j}} = 
-k_{t} \delta_{ij}^{\frac{1}{2}}\Bigg[(1 + \tilde{t} - \tilde{t}^{2})\pdv{t^{\beta}_{ij}}{\theta_{j}} 
+ (\tilde{t}^{\beta} - 2\tilde{t}\tilde{t}^{\beta})\pdv{t_{ij}}{\theta_{j}}\Bigg]
\end{equation}
From the above two equations, it is then understood that if $r_{ij}$ and $t_{ij}$ are known, the 
differential equations can be solved. When $\tilde{t}^{\beta}$ is negligible for all $\beta$, then 
$\tilde{t} \approx 0$, which then translates to 
$\pdv{\textnormal{F}^{t}_{ij}}{\theta_{j}} =  -k_{t} \delta^{\frac{1}{2}}\pdv{t^{\beta}_{ij}}{\theta_{j}} = 
-(-1)^{\beta}\,\textnormal{R}\delta^{\frac{1}{2}}_{ij}\,\frac{r_{ij}^{\alpha}}{r_{ij}}$, 
with $\alpha \ne \beta$, implying that even in the case of zero tangential displacement [hence, zero
tangential force], the above derivative can be finite.

The derivative of the normal force with respect to $r_{j}^{\alpha}$ follows:
\begin{equation}\label{eq28}
\begin{array} {lcl}
\pdv{{\textnormal{F}^{n}_{ij}}^{\beta}}{r^{\alpha}_{j}} & = & 
k_{n}\pdv{}{r^{\alpha}_{j}}\Bigg[\delta^{\frac{3}{2}}_{ij} \frac{r^{\alpha}_{ij}}{r_{ij}}\Bigg] \\
& = &  k_{n}\delta^{\frac{1}{2}}_{ij}\Bigg[-\Delta_{\alpha\beta}\frac{\delta_{ij}}{r_{ij}}
+ \frac{3}{2} \frac{r^{\alpha}_{ij}r^{\beta}_{ij}}{r^{2}_{ij}}  
+ \Big(\frac{\delta_{ij}}{r_{ij}}\Big)\frac{r^{\alpha}_{ij}r^{\beta}_{ij}}{r^{2}_{ij}}\Bigg] 
\end{array}
\end{equation}
where $\Delta_{\alpha\beta}$ is the Kronecker delta. 
Therefore, derivative of the total force is written as follow:
\begin{equation}\label{eq29}
\pdv{\textnormal{F}^{\beta}_{ij}}{r^{\alpha}_{j}} = 
\pdv{\textnormal{F}^{n\beta}_{ij}}{r^{\alpha}_{j}} + \pdv{\textnormal{F}^{t\beta}_{ij}}{r^{\alpha}_{j}}
\end{equation}
\begin{equation}\label{eq30}
\pdv{\textnormal{F}^{\beta}_{ij}}{\theta_{j}} = \pdv{\textnormal{F}^{t\beta}_{ij}}{\theta_{j}}
\end{equation}

The torque of a disk-$j$ due to tangential force 
$\textnormal{\textbf{F}}^{t}_{ij}$ is $\textbf{T}_{j} = 
-R(\hat{r}_{ij} \times \textbf{\textnormal{F}}^{t}_{ij}) \equiv R\tilde{\textnormal{\textbf{T}}}_{ij}$.
In a two-dimensional systems, $\tilde{\textnormal{\textbf{T}}}_{ij}$ has only $z$-component:
\begin{equation}\label{eq31}
\tilde{\textnormal{T}}^{z}_{ij} = -\Bigg[\Bigg(\frac{x_{ij}}{r_{ij}}\Bigg){\textnormal{F}^{t}_{ij}}^{y}
- \Bigg(\frac{y_{ij}}{r_{ij}}\Bigg){\textnormal{F}^{t}_{ij}}^{x} \Bigg]
\end{equation}
Therefore, the derivative of $\tilde{\textnormal{T}}^{z}_{ij}$ with respect to transnational coordinates 
$r^{\alpha}_{i}$ then becomes:
\begin{widetext}
\begin{equation}\label{eq32}
\pdv{\tilde{\textnormal{T}}^{z}_{ij}}{r^{\alpha}_{j}} = 
 \Bigg(\frac{\Delta_{\alpha x}}{r_{ij}} - \frac{x_{ij}r^{\alpha}_{ij}}{r^{3}_{ij}}\Bigg){\textnormal{F}^{t}_{ij}}^{y}
- \Bigg( \frac{x_{ij}}{r_{ij}}\Bigg)\pdv{{\textnormal{F}^{t}_{ij}}^{y}}{r^{\alpha}_{j}}
- \Bigg(\frac{\Delta_{\alpha y}}{r_{ij}} - \frac{y_{ij}r^{\alpha}_{ij}}{r^{3}_{ij}}\Bigg){\textnormal{F}^{t}_{ij}}^{x}
+ \Bigg( \frac{y_{ij}}{r_{ij}}\Bigg)\pdv{{\textnormal{F}^{t}_{ij}}^{x}}{r^{\alpha}_{j}}
\end{equation}
\end{widetext}
where $\Delta_{\alpha x}$ and, $\Delta_{\alpha y}$ is the Kronecker delta, i.e. 
$\Delta_{xx} = \Delta_{yy} = 1$ and zero otherwise. 
The derivative of the torque with respect to the $\theta_{i}$ would be:
\begin{equation}\label{eq33}
\pdv{\tilde{\textnormal{T}}^{z}_{ij}}{\theta_{j}} = -\Bigg[
\Bigg(\frac{x_{ij}}{r_{ij}}\Bigg)\pdv{\textnormal{F}^{y}_{ij}}{\theta_{j}}
-\Bigg(\frac{y_{ij}}{r_{ij}}\Bigg)\pdv{\textnormal{F}^{x}_{ij}}{\theta_{j}}
\Bigg]
\end{equation}
The above two differential equations can be solved using Eq.(\ref{eq26}), and Eq.(\ref{eq27}). If the 
tangential displacement $t^{\beta}_{ij}$ is negligible compared to the threshold $t^{\star}_{ij}$, 
i.e., $\tilde{t}^{\beta}_{ij} \approx 0$ for all $\beta$, this provides us with $\tilde{t} \approx 0$.
Therefore, $\pdv{\tilde{\textnormal{T}}^{z}_{ij}}{\theta_{j}} = k_{t}R\delta^{\frac{1}{2}}_{ij}$, which
means that even in the case of negligible/zero tangential displacement hence zero tangential force, 
the derivative of the torque is non-zero. 

{\bf Disk-wall interaction}: 
Following the Eq.(\ref{eq26}) for the disk-disk pair, we can write the derivative of the tangential 
force with respect $r^{\alpha}_{i}$ for the disk-wall pair as:
\begin{widetext}
\begin{equation}\label{eq34}
\begin{array}{l}
\pdv{{\textnormal{F}^{t}_{iw}}^{\beta}}{r^{\alpha}_{i}} = 
-k_{t} \pdv{}{r^{\alpha}_{i}}\Bigg[\delta_{iw}^{1/2}\Bigg(t^{\beta}_{iw} + \tilde{t}_{w}t^{\beta}_{iw} - 
\tilde{t}_{w}^{2}t^{\beta}_{iw}\Bigg)\Bigg]\\
= -\frac{1}{2} \delta^{-1}_{iw} \frac{r_{iw}^{\alpha}}{r_{iw}} {\textnormal{F}_{iw}^{t}}^{\beta} 
-\,\,k_{t}\delta^{1/2}_{iw}\Bigg[(1 + \tilde{t}_{w} - \tilde{t}_{w}^{2})\pdv{t^{\beta}_{iw}}{r^{\alpha}_{i}}
\,\,+\,\,(\tilde{t}_{w}^{\beta} - 2\tilde{t}_{w}\tilde{t}_{w}^{\beta})\pdv{t_{iw}}{r^{\alpha}_{i}} 
\,\,+\,\,(-\tilde{t}_{w}\tilde{t}_{w}^{\beta} + 2\tilde{t}_{w}^{2}\tilde{t}_{w}^{\beta})\pdv{t^{\star}_{iw}}{r^{\alpha}_{i}}\Bigg]
\end{array}
\end{equation}
\end{widetext}
Here, $\tilde{t}_{w} = \frac{t_{iw}}{t^{\star}_{iw}}$ and  
$\tilde{t}^{\beta}_{w} = \frac{t_{iw}^{\beta}}{t^{\star}_{iw}}$. 

Similarly, the derivative of tangential force with respect to $\theta_{i}$ (using the same notation as
above) can be found as:
\begin{equation}\label{eq35}
\pdv{\textnormal{F}^{t}_{iw}}{\theta_{i}} = 
-k_{t} \delta^{\frac{1}{2}}\Bigg[(1 + \tilde{t}_{w} - \tilde{t}_{w}^{2})\pdv{t^{\beta}_{iw}}{\theta_{i}} 
+ (\tilde{t}_{w}^{\beta} - 2\tilde{t}_{w}\tilde{t}_{w}^{\beta})\pdv{t_{iw}}{\theta_{i}}\Bigg]
\end{equation}

The derivative of the normal Hertzian force would be:
\begin{equation}\label{eq36}
\begin{array} {lcl}
\pdv{{\textnormal{F}^{n}_{iw}}^{\beta}}{r^{\alpha}_{i}} & = & 
k_{n}\pdv{}{r^{\alpha}_{i}}\Bigg[\delta^{\frac{3}{2}}_{iw} \frac{r^{\alpha}_{iw}}{r_{iw}}\Bigg] \\
& = &  k_{n}\delta^{\frac{1}{2}}_{iw}\Bigg[\Delta_{\alpha\beta}\frac{\delta_{iw}}{r_{iw}}
- \frac{3}{2} \frac{r^{\alpha}_{iw}r^{\beta}_{ij}}{r^{2}_{ij}} - 
\Big(\frac{\delta_{iw}}{r_{iw}}\Big)\frac{r^{\alpha}_{iw}r^{\beta}_{iw}}{r^{2}_{iw}}\Bigg] 
\end{array}
\end{equation}
where $\Delta_{\alpha\beta}$ is the Kronecker delta. In the above equation, one should note that
if $\textnormal{F}^{nx}_{iw} = 0$ (which is true in the current studies) 
then $\pdv{{\textnormal{F}^{n}_{iw}}^{\beta}}{r^{\alpha}_{i}} = 0$

Therefore, derivative of the total force is written as follow:
\begin{equation}\label{eq37}
\pdv{\textnormal{F}^{\beta}_{iw}}{r^{\alpha}_{i}} = 
\pdv{\textnormal{F}^{n\beta}_{iw}}{r^{\alpha}_{i}} + \pdv{\textnormal{F}^{t\beta}_{iw}}{r^{\alpha}_{i}}
\end{equation}
\begin{equation}\label{eq38}
\pdv{\textnormal{F}^{\beta}_{iw}}{\theta_{i}} = \pdv{\textnormal{F}^{t\beta}_{iw}}{\theta_{i}}
\end{equation}

The torque of a disk-$i$ due to tangential force 
$\textnormal{\textbf{F}}^{t}_{iw}$ is $\textbf{T}_{i} = 
-R(\hat{r}_{iw} \times \textbf{\textnormal{F}}^{t}_{iw}) \equiv R\tilde{\textnormal{\textbf{T}}}_{iw}$.
In a two-dimensional systems, $\tilde{\textnormal{\textbf{T}}}_{iw}$ has only $z$-component:
\begin{equation}\label{eq39}
\tilde{\textnormal{T}}^{z}_{iw} = -\Bigg[\Bigg(\frac{x_{iw}}{r_{iw}}\Bigg){\textnormal{F}^{t}_{iw}}^{y}
- \Bigg(\frac{y_{iw}}{r_{iw}}\Bigg){\textnormal{F}^{t}_{iw}}^{x} \Bigg]
\end{equation}
Therefore, the derivative of $\tilde{\textnormal{T}}^{z}_{iw}$ with respect to transnational coordinates 
$r^{\alpha}_{i}$ then becomes:
\begin{widetext}
\begin{equation}\label{eq40}
\pdv{\tilde{\textnormal{T}}^{z}_{iw}}{r^{\alpha}_{i}} = 
- \Bigg(\frac{\Delta_{\alpha x}}{r_{iw}} - \frac{x_{iw}r^{\alpha}_{iw}}{r^{3}_{iw}}\Bigg){\textnormal{F}^{t}_{iw}}^{y}
- \Bigg( \frac{x_{iw}}{r_{iw}}\Bigg)\pdv{{\textnormal{F}^{t}_{iw}}^{y}}{r^{\alpha}_{iw}}
+ \Bigg(\frac{\Delta_{\alpha y}}{r_{iw}} - \frac{y_{iw}r^{\alpha}_{iw}}{r^{3}_{iw}}\Bigg){\textnormal{F}^{t}_{iw}}^{x}
+ \Bigg( \frac{y_{iw}}{r_{iw}}\Bigg)\pdv{{\textnormal{F}^{t}_{iw}}^{x}}{r^{\alpha}_{iw}}
\end{equation}
\end{widetext}
where $\Delta_{\alpha x}$ and, $\Delta_{\alpha y}$ is the Kronecker delta, i.e. 
$\Delta_{xx} = \Delta_{yy} = 1$ and zero otherwise. 
The derivative of the torque with respect to the $\theta_{i}$ would be:
\begin{equation}\label{eq41}
\pdv{\tilde{\textnormal{T}}^{z}_{iw}}{\theta_{i}} = -\Bigg[
\Bigg(\frac{x_{iw}}{r_{iw}}\Bigg)\pdv{\textnormal{F}^{y}_{iw}}{\theta_{i}}
-\Bigg(\frac{y_{iw}}{r_{iw}}\Bigg)\pdv{\textnormal{F}^{x}_{iw}}{\theta_{i}}
\Bigg]
\end{equation}

{\bf Expressions for the different parts of the Jacobian}:
The dimension of Jacobian operator $\mathbb{J}$ is force over length. To be consistent with the dimension, we 
redefine the torque T and rotational coordinate $\theta_{i}$ as:
\begin{equation}\label{eq42}
\tilde{\textnormal{T}}_{i} = \frac{\textnormal{T}_{i}}{\textnormal{R}}  
\quad \textnormal{and} \quad
\tilde{\theta_{i}} = \textnormal{R}\theta_{i}
\end{equation}
In addition, the dynamical matrix has a contribution from the moment of inertia 
$\textnormal{I}_{i} = \textnormal{I}_{0}m_{i}\textnormal{R}^{2}$ since 
$\Delta \boldsymbol{\omega}_{i} = \frac{\textbf{T}_{i}}{\textnormal{I}_{i}\Delta t}$. In our 
calculation, we assume that mass $m_{i}$ and $\textnormal{I}_{0}$ both are 
one. The remaining contribution of $\textnormal{I}_{i}$ i.e. $\textnormal{R}^{2}$, is taken care of by
re-scaling the torque and angular displacement as $\tilde{\textnormal{T}}_{i}$ and 
$\tilde{\theta}_{i}$. For $\textnormal{I}_{i} \ne 1$, the contribution of
$\textnormal{I}_{i}$ can be correctly anticipated if we rewrite Eq.(\ref{eq12}) as below:
\begin{equation}\label{eq43}
\frac{\textnormal{d}\textbf{t}_{ij}}{\textnormal{d}t} = \textbf{v}_{ij} - \textbf{v}_{ij}^{n} 
+ \frac{1}{\textnormal{I}_{0}}\hat{r}_{ij} \times R(\boldsymbol{\omega}_{i} + \boldsymbol{\omega}_{j})
\end{equation}

The Jacobian operator $\mathbb{J} = \Big|\Big|\pdv{\textbf{F}_{i}}{\textbf{Q}_{j}}\Big|\Big|_{\bf Q_{{\bf 0}}}$
can be divided into four parts, given as: (i) Derivative of the force $F^{\beta}_{i}$ with respect to 
transnational coordinates $r^{\alpha}_{j}$ (ii) Derivative of the force $F^{\beta}_{i}$ with respect to 
angular displacements $\theta_{j}$ (iii) Derivative of the torque $T^{\beta}_{i}$ with respect to 
transnational coordinates $r^{\alpha}_{j}$ (iv) Derivative of the torque $F^{\beta}_{i}$ with respect 
to angular displacements $\theta_{j}$. 
Total force on disk $i$ is given by:
\begin{equation}\label{eq44}
\textbf{F}_{i} = \sum_{j = 1, i \ne j}^{N} \textbf{F}_{ij} + \textbf{F}_{iw} + \textbf{F}_{i}^{y} 
+ \textbf{F}_{i}^{x}
\end{equation}
where,
\begin{enumerate}[label={(\roman*)}]
\item $\sum_{j}^{N} \textbf{F}_{ij}$ = total force on the disk due to other disk in contact
\item $\textbf{F}_{iw}$ = total force on the disk due to wall
\item $\textbf{F}_{i}^{y}$ = constant force on all the disks pushing the disks down 
(-$\hat{y}$ direction)
\item $\textbf{F}_{i}^{x}$ = constant force on disk ``$i = 1$'' pushing the disk in horizontally
(+$\hat{x}$ direction). $\textbf{F}_{i}^{x}$ = 0 for $i \ne 1$ 
\end{enumerate}
Differentiating both sides with respect to $\bf{r}_{j}$\\
$$
\pdv{\textbf{F}_{i}}{\textbf{r}_{j}} = 
\frac{\partial}{\partial\textbf{r}_{j}}\sum_{j=1,j\ne i}^{N} \textbf{F}_{ij}
+ \frac{\partial}{\partial\textbf{r}_{j}}\textbf{F}_{iw}
+ \cancelto{0}{\frac{\partial}{\partial\textbf{r}_{j}}\textbf{F}_{i}^{y}}
+ \cancelto{0}{\frac{\partial}{\partial\textbf{r}_{j}}\textbf{F}_{i}^{x}}
$$ 
\begin{equation}\label{eq45}
\pdv{\textbf{F}_{i}}{\textbf{r}_{j}} = 
\frac{\partial}{\partial\textbf{r}_{j}}\sum_{j=1,j\ne i}^{N} \textbf{F}_{ij}
+ \frac{\partial}{\partial\textbf{r}_{j}}\textbf{F}_{iw}
\end{equation}
Using tensor notations, we get
\begin{equation}\label{eq46}
\pdv{F^{\beta}_{i}}{r^{\alpha}_{j}} = 
\frac{\partial}{\partial r^{\alpha}_{j}}\sum_{j=1,j\ne i}^{N} F^{\beta}_{ij}
+ \frac{\partial F^{\beta}_{iw}}{\partial r^{\alpha}_{j}}
\end{equation}
The Jacobian matrix can be divided into four parts. Details of each part is provided below:\\\\
{\bf First part}: Derivative of forces with respect to positions of particles follows as:
\begin{equation}\label{eq47}
\pdv{F^{\beta}_{i}}{r^{\alpha}_{j}} = 
\frac{\partial}{\partial r^{\alpha}_{j}}\sum_{j=1,j\ne i}^{N} F^{\beta}_{ij}
+ \cancelto{0}{\frac{\partial F^{\beta}_{iw}}{\partial r^{\alpha}_{j}}}
= \frac{\partial}{\partial r^{\alpha}_{j}}\sum_{j=1,j\ne i}^{N} F^{\beta}_{ij}
\end{equation}
\begin{equation}\label{eq48}
J^{\alpha \beta}_{ij} = \pdv{F^{\beta}_{i}}{r^{\alpha}_{j}} = \frac{\partial F^{\beta}_{ij}}{\partial r^{\alpha}_{j}}
;\hspace{1cm}\,for\,\, j \ne i
\end{equation}
\begin{equation}\label{eq49}
\pdv{F^{\beta}_{i}}{r^{\alpha}_{i}} = \frac{\partial}{\partial r^{\alpha}_{i}}\sum_{j=1,j\ne i}^{N} F^{\beta}_{ij}
+ \frac{\partial F^{\beta}_{iw}}{\partial r^{\alpha}_{i}}
\end{equation}
\begin{equation}\label{eq50}
\pdv{F^{\beta}_{i}}{r^{\alpha}_{i}} = -\sum_{j=1,j\ne i}^{N}\frac{\partial F^{\beta}_{ij}}{\partial r^{\alpha}_{j}} 
+ \frac{\partial F^{\beta}_{iw}}{\partial r^{\alpha}_{i}}
;\hspace{1cm}\,for\,\, j = i
\end{equation}
which then gives us
\begin{equation}\label{eq51}
\begin{aligned}
J^{\alpha \beta}_{ij} &= \frac{\partial F^{\beta}_{ij}}{\partial r^{\alpha}_{j}}\\\\
J^{\alpha \beta}_{ii} &= -\sum_{j=1,j\ne i}^{N}\frac{\partial F^{\beta}_{ij}}{\partial r^{\alpha}_{j}} 
+ \frac{\partial F^{\beta}_{iw}}{\partial r^{\alpha}_{i}}
\end{aligned}
\end{equation}
It should be noted that in the derivative of Hertzian normal force for the grain-wall interactions, 
following constrained must be applied:
\begin{widetext}
\begin{equation}\label{eq52}
\pdv{F^{n\beta}_{iw}}{r^{\alpha}_{i}} = k_{n}\delta^{1/2}
\Big[\Delta_{\alpha\beta}\frac{\delta_{ij}}{r_{ij}} -\frac{3}{2}\frac{r_{ij}^{\alpha}r_{ij}^{\beta}}{r_{ij}^{2}} 
- \Big(\frac{\delta_{ij}}{r_{ij}}\Big)\Big(\frac{r_{ij}^{\alpha}r_{ij}^{\beta}}{r_{ij}^{2}}\Big)\Big]
= 0,\,\textnormal{if}\,r_{iw}^{n\beta}\,=\,0\,\textnormal{i.e.}\,\,F_{iw}^{n\beta}\,=\,0
\end{equation}
\end{widetext}

{\bf Second part}: Derivative of forces with respect to rotational coordinates of particles:\\
\begin{equation}\label{eq53}
\pdv{F^{\beta}_{i}}{\tilde{\theta}_{j}} = 
\frac{\partial}{\partial \tilde{\theta}_{j}}\sum_{j=1,j\ne i}^{N} F^{\beta}_{ij}
+ \cancelto{0}{\frac{\partial F^{\beta}_{iw}}{\partial \tilde{\theta}_{j}}}
= \frac{\partial}{\partial \tilde{\theta}_{j}}\sum_{j=1,j\ne i}^{N} F^{\beta}_{ij}
\end{equation}
\begin{equation}\label{eq54}
\pdv{F^{\beta}_{i}}{\tilde{\theta}_{j}} = \frac{\partial F^{\beta}_{ij}}{\partial \tilde{\theta}_{j}}
;\hspace{1cm}\,for\,\, j \ne i
\end{equation}
\begin{equation}\label{eq55}
\pdv{F^{\beta}_{i}}{\tilde{\theta}_{i}} = \frac{\partial}{\partial \tilde{\theta}_{i}}\sum_{j=1,j\ne i}^{N} F^{\beta}_{ij}
+ \frac{\partial F^{\beta}_{iw}}{\partial \tilde{\theta}_{i}}
\end{equation}
\begin{equation}\label{eq56}
\pdv{F^{\beta}_{i}}{\tilde{\theta}_{i}} = \sum_{j=1,j\ne i}^{N}\frac{\partial F^{\beta}_{ij}}{\partial \tilde{\theta}_{i}} 
+ \frac{\partial F^{\beta}_{iw}}{\partial \tilde{\theta}_{i}}
;\hspace{1cm}\,\textnormal{for}\,\, j = i
\end{equation}
which gives
\begin{equation}\label{eq57}
\begin{aligned}
J^{\beta}_{ij} &= \frac{\partial F^{\beta}_{ij}}{\partial \tilde{\theta}_{i}}\\\\
J^{\beta}_{ii} &= \sum_{j=1,j\ne i}^{N}\frac{\partial F^{\beta}_{ij}}{\partial \tilde{\theta}_{i}} 
+ \frac{\partial F^{\beta}_{iw}}{\partial \tilde{\theta}_{i}}
\end{aligned}
\end{equation}
{\bf Third part}: Derivative of torques with respect to positions of particles:\\
\begin{equation}\label{eq58}
\pdv{\tilde{T}^{z}_{i}}{r^{\alpha}_{j}} = 
\frac{\partial}{\partial r^{\alpha}_{j}}\sum_{j=1,j\ne i}^{N} \tilde{T}^{z}_{ij}
+ \cancelto{0}{\frac{\partial \tilde{T}^{z}_{iw}}{\partial r^{\alpha}_{j}}}
= \frac{\partial}{\partial r^{\alpha}_{j}}\sum_{j=1,j\ne i}^{N} T^{z}_{ij}
\end{equation}
\begin{equation}\label{eq59}
\pdv{\tilde{T}^{z}_{i}}{r^{\alpha}_{j}} = \frac{\partial \tilde{T}^{z}_{ij}}{\partial r^{\alpha}_{j}}
;\hspace{1cm}\,\textnormal{for}\,\, j \ne i
\end{equation}
\begin{equation}\label{eq60}
\pdv{\tilde{T}^{z}_{i}}{r^{\alpha}_{i}} = \frac{\partial}{\partial r^{\alpha}_{i}}\sum_{j=1,j\ne i}^{N} \tilde{T}^{z}_{ij}
+ \frac{\partial \tilde{T}^{z}_{iw}}{\partial r^{\alpha}_{i}}
\end{equation}
\begin{equation}\label{eq61}
\pdv{\tilde{T}^{z}_{i}}{r^{\alpha}_{i}} = -\sum_{j=1,j\ne i}^{N}\frac{\partial \tilde{T}^{z}_{ij}}{\partial r^{\alpha}_{j}} 
+ \frac{\partial \tilde{T}^{z}_{iw}}{\partial r^{\alpha}_{i}}
;\hspace{1cm}\,for\,\, j = i
\end{equation}
which gives
\begin{equation}\label{eq62}
\begin{aligned}
J^{\beta}_{ij} &= \frac{\partial \tilde{T}^{z}_{ij}}{\partial r^{\beta}_{j}}\\\\
J^{\beta}_{ii} &=  -\sum_{j=1,j\ne i}^{N}\frac{\partial \tilde{T}^{z}_{ij}}{\partial r^{\beta}_{j}} 
+ \frac{\partial \tilde{T}^{z}_{iw}}{\partial r^{\beta}_{i}}
\end{aligned}
\end{equation}
{\bf Fourth part}: Derivative of torques with respect to rotational coordinates of particles:
\begin{equation}\label{eq63}
\pdv{\tilde{T}^{z}_{i}}{\tilde{\theta}_{j}} = 
\frac{\partial}{\partial \tilde{\theta}_{j}}\sum_{j=1,j\ne i}^{N} \tilde{T}^{z}_{ij}
+ \cancelto{0}{\frac{\partial \tilde{T}^{z}_{iw}}{\partial \tilde{\theta}_{j}}}
= \frac{\partial}{\partial \tilde{\theta}_{j}}\sum_{j=1,j\ne i}^{N} \tilde{T}^{z}_{ij}
\end{equation}
\begin{equation}\label{eq64}
\pdv{\tilde{T}^{z}_{i}}{\tilde{\theta}_{j}} = \frac{\partial \tilde{T}^{z}_{ij}}{\partial \tilde{\theta}_{j}}
;\hspace{1cm}\,for\,\, j \ne i
\end{equation}
\begin{equation}\label{eq65}
\pdv{\tilde{T}^{z}_{i}}{\tilde{\theta}_{i}} = \frac{\partial}{\partial \tilde{\theta}_{i}}\sum_{j=1,j\ne i}^{N} \tilde{T}^{z}_{ij}
+ \frac{\partial \tilde{T}^{z}_{iw}}{\partial \tilde{\theta}_{i}}
\end{equation}
\begin{equation}\label{eq66}
\pdv{\tilde{T}^{z}_{i}}{\tilde{\theta}_{i}} = \sum_{j=1,j\ne i}^{N}\frac{\partial \tilde{T}^{z}_{ij}}{\partial \tilde{\theta}_{j}} 
+ \frac{\partial \tilde{T}^{z}_{iw}}{\partial \tilde{\theta}_{i}}
;\hspace{1cm}\,for\,\, j = i
\end{equation}
which gives
\begin{equation}\label{eq67}
\begin{aligned}
J_{ij} & = \frac{\partial \tilde{T}^{z}_{ij}}{\partial \tilde{\theta}_{j}}\\\\
J_{ii} & =  \sum_{j=1,j\ne i}^{N}\frac{\partial \tilde{T}^{z}_{ij}}{\partial \tilde{\theta}_{j}} 
+ \frac{\partial \tilde{T}^{z}_{iw}}{\partial \tilde{\theta}_{i}}
\end{aligned}
\end{equation}

{\bf An example of the Jacobian matrix for a 2 disks system}:
For each disk the total degrees of freedom are 3, i.e. two (d) in $x$ and $y$ and the third one is the 
$\theta$ rotations. Hence, the dimension of the Jacobian matrix is $(d + 1)N\times(d + 1)N$. Here we 
show the arrangement of the elements of the Jacobian matrix for a two disk system. 
\begin{center}
\begin{math}
J = 
\begin{bmatrix}
\frac{\partial F_{1}^{x}}{\partial x_{1}}  & \frac{\partial F_{1}^{x}}{\partial x_{2}} 
& \frac{\partial F_{1}^{x}}{\partial y_{1}}
& \frac{\partial F_{1}^{x}}{\partial y_{2}} & \frac{\partial F_{1}^{x}}{\partial \theta_{1}}
& \frac{\partial F_{1}^{x}}{\partial \theta_{2}} \\\\
\frac{\partial F_{2}^{x}}{\partial x_{1}}  & \frac{\partial F_{2}^{x}}{\partial x_{2}} 
& \frac{\partial F_{2}^{x}}{\partial y_{1}}
& \frac{\partial F_{2}^{x}}{\partial y_{2}} & \frac{\partial F_{2}^{x}}{\partial \theta_{1}}
& \frac{\partial F_{2}^{x}}{\partial \theta_{2}} \\\\
\frac{\partial F_{1}^{y}}{\partial x_{1}}  & \frac{\partial F_{1}^{y}}{\partial x_{2}} 
& \frac{\partial F_{1}^{y}}{\partial y_{1}}
& \frac{\partial F_{1}^{y}}{\partial y_{2}} & \frac{\partial F_{1}^{y}}{\partial \theta_{1}}
& \frac{\partial F_{1}^{y}}{\partial \theta_{2}} \\\\
\frac{\partial F_{2}^{y}}{\partial x_{1}}  & \frac{\partial F_{2}^{y}}{\partial x_{2}} 
& \frac{\partial F_{2}^{y}}{\partial y_{1}}
& \frac{\partial F_{2}^{y}}{\partial y_{2}} & \frac{\partial F_{2}^{y}}{\partial \theta_{1}}
& \frac{\partial F_{2}^{y}}{\partial \theta_{2}} \\\\
\frac{\partial \tilde{T}_{1}}{\partial x_{1}}  & \frac{\partial \tilde{T}_{1}}{\partial x_{2}} 
& \frac{\partial \tilde{T}_{1}}{\partial y_{1}}
& \frac{\partial \tilde{T}_{1}}{\partial y_{2}} & \frac{\partial \tilde{T}_{1}}{\partial \tilde{\theta}_{1}}
& \frac{\partial \tilde{T}_{1}}{\partial \tilde{\theta}_{2}} \\\\
\frac{\partial \tilde{T}_{2}}{\partial x_{1}}  & \frac{\partial \tilde{T}_{2}}{\partial x_{2}} 
& \frac{\partial \tilde{T}_{2}}{\partial y_{1}}
& \frac{\partial \tilde{T}_{2}}{\partial y_{2}} & \frac{\partial \tilde{T}_{2}}{\partial \tilde{\theta}_{1}}
& \frac{\partial \tilde{T}_{2}}{\partial \tilde{\theta}_{2}} \\\\
\end{bmatrix}
\end{math}
\end{center}
%